\documentclass[epj]{svjour}

\def\be{\begin{equation}} 
\def\ee{\end{equation}}   

\RequirePackage{graphicx}

\usepackage{cite} 
\usepackage{amsmath}
\usepackage{hyperref}
\usepackage{cite}
\usepackage{amsmath,amssymb}
\usepackage{xcolor}
\usepackage[utf8]{inputenc}

\usepackage{graphics}

\usepackage{cite} 
\usepackage{amsmath}
\usepackage{hyperref}
\usepackage{cite}
\usepackage{amsmath,amssymb}
\usepackage{xcolor}

\begin{document}

\title{Radial oscillations and tidal Love numbers of dark energy stars}

\author{
Grigoris Panotopoulos \inst{1} 
\thanks{E-mail: \href{mailto:grigorios.panotopoulos@tecnico.ulisboa.pt}{\nolinkurl{grigorios.panotopoulos@tecnico.ulisboa.pt}} }
\and
\'Angel Rinc\'on \inst{2}
\thanks{E-mail: \href{mailto:angel.rincon@pucv.cl}{\nolinkurl{angel.rincon@pucv.cl}} }
\and
Il{\'i}dio Lopes \inst{1} 
\thanks{E-mail: \href{mailto:ilidio.lopes@tecnico.ulisboa.pt}{\nolinkurl{ilidio.lopes@tecnico.ulisboa.pt}} }
}

\institute{ 
Centro de Astrof\'{\i}sica e Gravita{\c c}{\~a}o, Departamento de F{\'i}sica, Instituto Superior T\'ecnico-IST, \\
Universidade de Lisboa-UL, Av. Rovisco Pais, 1049-001 Lisboa, Portugal.
\and
Instituto de F{\'i}sica, Pontificia Universidad Cat{\'o}lica de Valpara{\'i}so, Avenida Brasil 2950, Casilla 4059, Valpara{\'i}so, Chile.
}

\date{Received: date / Revised version: date}

\abstract{
We investigate the properties of relativistic stars made of dark energy. 
We model stellar structure assuming i) isotropic perfect fluid and ii) a dark energy inspired equation of state, the generalized equation of state of Chaplygin gas, as we will be calling it. The mass-to-radius profiles, the tidal Love numbers as well as the ten lowest radial oscillation modes are computed. Causality, stability and energy conditions are also discussed.
\PACS{
      {PACS-key}{discribing text of that key}   \and
      {PACS-key}{discribing text of that key}
     } 
} 

\maketitle

\section{Introduction}\label{Intro}

The origin and nature of dark energy (DE), the fluid component that currently accelerates the Universe \cite{SN1,SN2,turner}, is one of the biggest mysteries and challenges in modern theoretical Cosmology. It is well-known that Einstein's General Relativity (GR) \cite{GR} with radiation and matter only cannot lead to accelerating solutions. Since the discovery of the current cosmic acceleration, a positive cosmological constant \cite{einstein}, despite its simplicity, has become the concordance cosmological model in a very 
good agreement with a vast amount of modern observational data.

\smallskip

The $\Lambda$CDM model (concordance model), based on cold dark matter and a positive cosmological constant, however, suffers from the cosmological constant problem \cite{zeldovich,weinberg}. Furthermore, regarding the value of the Hubble constant $H_0$, there is nowadays a tension between high red-shift CMB data and local measurements at low red-shift data, see e.g. \cite{tension,2019Colin,tension1,tension2,tension3}. The value of the Hubble constant extracted by the PLANCK Collaboration \cite{planck1,planck2}, $H_0 = (67-68)~\text{km/(Mpc  sec)}$, is found to be lower than the value obtained by local measurements, $H_0 = (73-74)~\text{km/(Mpc sec)}$ \cite{hubble,recent}. This tension might call for new physics \cite{newphysics,Alvarez:2020xmk}.

\smallskip

For those reasons, several alternatives to the $\Lambda$CDM model have been proposed and studied over the years. Generically speaking, all dark energy (DE) models fall into two broad classes, namely either a modified theory of gravity is assumed, providing correction terms to GR at cosmological scales, or a new dynamical degree of freedom with an equation of state  parameter $w < -1/3$ must be introduced. In the first class of models (geometrical DE) one finds for instance $f(R)$ theories of gravity \cite{mod1,mod2,HS,starobinsky}, brane-world models \cite{langlois,maartens,dgp} and Scalar-Tensor theories of gravity \cite{BD1,BD2,leandros,PR}, while in the second class (dynamical DE) one finds models such as quintessence \cite{DE1}, phantom \cite{DE2}, quintom \cite{DE3}, tachyonic \cite{DE4} or k-essence \cite{DE5}. For an excellent review on the dynamics of dark energy see e.g. \cite{copeland}.

\smallskip

Compact objects \cite{textbook}, such as neutron stars \cite{NSreview}, are characterized by ultra high matter densities, and strong gravitational fields. As matter becomes relativistic under those extreme conditions, the Newtonian description is inadequate, and therefore a fully relativistic description, based on the highly non-linear field equations of GR, is required. Stellar structure, modelling of relativistic compact objects, and obtaining exact analytical solutions to Einstein's field equations is both challenging and interesting, and it has been keeping researchers busy for decades now. 

\smallskip

Strange quark stars \cite{SS1,SS2,SS3,SS4,SS5,SS6}, although less conventional than neutron stars, are well motivated compact objects for at least two reasons. The first one is that they may explain some puzzling super-luminous supernovae \cite{SL1,SL2}, which occur in about one out of every 1000 supernovae explosions, and which are more than 100 times more luminous than regular supernovae. The second reason in favour of strange quark stars is the following: Despite the fact that as of today they comprise a speculative class of relativistic compact objects, quark stars cannot conclusively be ruled out yet. Indeed, there are some claims in the literature that there are currently some observed astronomical objects with peculiar features, such as small radii, that cannot be explained adopting the known hadronic equations-of-state (EoS) for neutron stars. For the relevant discussion see for instance \cite{cand1,cand2,cand3}, and also Table 5 of \cite{weber} and references therein.

\smallskip

Beyond the collisionless dark matter paradigm, self-interacting dark matter has been proposed as an attractive solution to the dark matter crisis at galactic scales \cite{Tulin}. In this scenario one can imagine relativistic stars made entirely of self-interacting dark matter, see e.g. \cite{DMS1,DMS2,DMS3}. In a similar way, given that the current cosmic acceleration calls for dark energy, very recently a couple of works appeared in the literature, where the authors entertain the possibility that stars made of dark energy or more generically exotic matter just might exist \cite{exotic,paperbase}.
In the present work, we propose to study non-rotating dark energy stars with isotropic matter assuming a generalized  equation-of-state of the form $p = -B^2/\rho + A^2\rho$ (A and B are constants). A simplified version of this, known as a Chaplygin equation-of-state, was introduced in Cosmology long time ago to unify the description of non-relativistic matter and the cosmological constant \cite{Chaplygin1,Chaplygin2,Chaplygin3}.
Such a generic equation-of-state is originated by a viscose matter, that when considered in a cosmological context gives rise to the unification of dark matter and dark energy \cite{Chaplygin4}.

\smallskip

The inspiral and relativistic collision of two compact objects in a binary system, and the gravitational wave signal emitted during the process, contain a wealth of information on the nature of the colliding bodies. The impact of an assumed EoS on the signals emitted during a binary coalescence is determined by adiabatic tidal interactions, characterized by a handful of coefficients, known in the literature as the tidal deformability, and the corresponding tidal Love numbers \cite{Love1,Love2}. What is more, stellar seismology has been a successful tool to infer the internal properties of the Sun and other similar stars, allowing astronomers to have a detail characterization of the microphysics and fluid dynamics processes occurring in their interiors, for instance, nuclear reactions, equation-of-state, differential rotation rate and meridional circulation \cite{pulsating1a,pulsating1aa}. Moreover, those techniques are now being extended to the study of the internal structure of compact stars like white dwarfs and neutron stars \cite{pulsating1c,pulsating1d,pulsating1e}. 
Those new methods of diagnostic provide a robust way to look for hints of physics inside stars, such as the existence of dark matter \cite{pulsating1b,pulsating2c} or alternative theories of gravity \cite{pulsating2a,pulsating2b}.
Therefore, in this work by computing the oscillations of these new class of stars, i.e., the frequencies and eigenfunctions of the radial mode oscillations, e.g. \cite{stellarpul1,stellarpul2},
we can learn about their composition, and their equation-of-state of the strongly interacting matter, since the precise values of these frequency modes are very sensitive to the underlying physics and internal structure of the star, see e.g. \cite{pulsating1, pulsating2, pulsating3, pulsating4, pulsating5, pulsating6, pulsating7, pulsating8, pulsating9} and references therein.

\smallskip

In the present work we compute the tidal Lover numbers and the radial oscillation frequencies for exotic stars with a dark energy inspired non-linear equation of state. The plan of the manuscript is the following: In the next section we briefly review the hydrostatic equilibrium of relativistic stars, while in section 3 we summarize the theory of radial oscillations as well as tidal Love numbers. In the fourth section we integrate the structure equations numerically, and we discuss the properties of the stars. Finally, we close our work with some concluding remarks in section 5. We adopt the mostly positive metric signature, $(-,+,+,+)$, and we work in geometrical units where the speed of light in vacuum as well as Newton's constant are set to unity, $c=1=G$.

\section{Hydrostatic equilibrium of relativistic stars}

To obtain interior solutions describing hydrostatic equilibrium of relativistic stars within GR, one needs to integrate the Tolman-Oppenheimer-Volkoff equations \cite{Tolman,OV}
\begin{eqnarray}
m'(r) & = & 4 \pi r^2 \rho(r) \\
p'(r) & = & - [\rho(r) + p(r)] \: \frac{m(r)+4 \pi r^3 p(r)}{r^2 (1-2m(r)/r)} \\
\nu'(r) & = & 2 \frac{m(r)+4 \pi r^3 p(r)}{r^2 (1-2m(r)/r)} 
\end{eqnarray}
where $p(r)$ and $\rho(r)$ are the pressure and the energy density of the matter content, while $\nu(r)$ and $m(r)$ are the components of the metric tensor assuming static, spherically symmetric solutions in Schwarzschild coordinates, $(t,r,\theta,\phi)$
\begin{equation}
ds^2 = -e^{\nu} dt^2 + \frac{1}{1-2m(r)/r} dr^2 + r^2 (d \mathrm{\theta^2} + \mathrm{sin^2 \theta \: d \phi^2})
\end{equation}
and all unknown quantities depend on the radial coordinate $r$ only.
Those equations are integrated imposing at the centre of the star the initial conditions \cite{pulsating9}
\begin{eqnarray}
m(0) & = & 0  \\
p(0) & = & p_c
\end{eqnarray}
with $p_c$ being the central pressure. Finally, upon matching with the exterior Schwarzschild vacuum solution
at the surface of the stars, the following conditions must be satisfied \cite{pulsating9}
\begin{eqnarray}
p(R) & = & 0  \\
m(R) & = & M \\
e^{\nu(R)} & = & 1 - \frac{2M}{R}
\end{eqnarray}
where $M$ is the mass of the star, while $R$ is its radius.

Finally, to integrate the structure equations we must specify the sources, i.e. the equation of state obeyed by matter. Following \cite{paperbase} we consider the dark energy inspired equation of state of the form
\begin{equation}
p = -\frac{B^2}{\rho} + A^2 \rho
\end{equation}
where $A,B$ are positive constants. In particular, $A$ is dimensionless, while $B$ has dimensions of energy density. Here we shall consider three concrete models as follows:
\begin{equation}
A = \sqrt{0.4}, \ \quad B=0.23 \times 10^{-3}/km^2 \; \; \; (\textrm{model A or 1})
\end{equation}
\begin{equation}
A = \sqrt{0.425}, \ \quad B=0.215 \times 10^{-3}/km^2 \; \; \; (\textrm{model B or 2})
\end{equation}
\begin{equation}
A = \sqrt{0.45}, \ \quad B=0.2 \times 10^{-3}/km^2 \; \; \; (\textrm{model C or 3}).
\end{equation}
Note that when the pressure vanishes at the surface of the star, the energy density 
takes the surface value $\rho_s = B/A$.


\begin{table}
\begin{center}
\begin{tabular}{l | l l l l l}
 & \multicolumn{4}{c}{{\sc Properties of six fiducial stars}} \\
Stars & $M (M_{\odot})$  & $R$ (km) & $C=M/R$ & $\omega_0$ (kHz) \\
\hline
\hline
 $A_1$  & 1.90  & 11.69 & 0.240 &  12.574  \\
 $A_2$  & 2.13  & 11.98 & 0.263 &  12.832  \\
 $B_1$  & 1.90  & 12.17 & 0.231 &  11.839  \\
 $B_2$  & 2.13  & 12.52 & 0.252 &  12.012  \\
 $C_1$  & 1.90  & 12.66 & 0.222 &  11.157  \\
 $C_2$  & 2.13  & 13.05 & 0.241 &  11.281  \\
\end{tabular}
\caption{
Six fiducial stars are considered in this work from the three models ($A$, $B$ and $C$).
The fundamental frequency $\omega_0$ is given by the expression $\omega_0=\sqrt{M/R^3}$.
}
\label{tab:1set}
\end{center}
\end{table}


\begin{table}
\begin{tabular}{l | l l l l l l}
& \multicolumn{6}{c}{{\sc Radial oscillation modes for different stars}} \\
$n$ & $A_{1}$  & $A_{2}$  & $B_{1}$  & $B_{2}$  & $C_{1}$  & $C_{2}$  \\
\hline
\hline
0  & 4.686    & 3.848   & 5.016    & 4.255  & 5.298   & 4.585  \\
1  & 10.913   & 9.332   & 11.446   & 9.986  & 11.905   & 10.523  \\
2  & 16.769  & 14.414  & 17.539   & 15.359  & 18.203  & 16.137  \\
3  & 22.543  & 19.410  & 23.557   & 20.655  & 24.430  & 21.679  \\
4  & 28.285  & 24.373  & 29.545   & 25.920  & 30.629  & 27.193  \\
5  & 34.011  & 29.319  & 35.518   & 31.169  & 36.815  & 32.693  \\
6  & 39.728  & 34.256  & 41.482   & 36.411  & 42.993  & 38.185  \\
7  & 45.440  & 39.188  & 47.442   & 41.647  & 49.167  & 43.672  \\
8  & 51.147  & 44.115  & 53.398   & 46.879  & 55.336  & 49.156  \\
9  & 56.852  & 49.039  & 59.352   & 52.109  & 61.504  & 54.636  \\
\end{tabular}
\caption{Frequencies $\nu_n$ in $kHz$ for the radial modes of
 six fiducial stars considered here (see Table~\ref{tab:1set}). $n$ is the  order of the radial mode.}
\label{tab:2set}
\centering
\end{table}


\section{Equations for the perturbations}

\subsection{Tidal Love numbers}

The theory of tidal Love numbers may be found for instance in \cite{Hinderer,Damour,Lattimer}. The deformabilities $\lambda, \Lambda$ may be computed in terms of the tidal Love number $k$, and they are defined by
\begin{equation}
\lambda \equiv \frac{2}{3} k R^5
\end{equation}
\begin{equation}
\Lambda \equiv \frac{k}{C^5}
\end{equation}
with $C=M/R$ being the factor of compactness of the star.

In terms of $C$, the tidal Love number is given by \cite{Hinderer,Lattimer}
\begin{equation}
k = \frac{8C^5}{5} \: \frac{K_{o}}{3  \:K_{o} \: \ln(1-2C) + P_5(C)} 
\label{elove}
\end{equation}
\begin{equation}
K_{o}=(1-2C)^2 \: [2 C (y_R-1)-y_R+2]
\end{equation}
\begin{equation}
y_R \equiv y(r=R)
\end{equation}
where $P_5(C)$ is a fifth order polynomial given by
\begin{align}
\begin{split}
P_5(C) = & 2 C \: [4 C^4 (y_R+1) + 2 C^3 (3 y_R-2) \ + 
\\
&
2 C^2 (13-11 y_R) + 3 C (5 y_R-8) -3 y_R + 6] 
\end{split}
\end{align}
while the function $y(r)$ satisfies a Riccati differential equation, 
see e.g \cite{Lattimer} for more details.

\subsection{Radial oscillations}

The equations for radial perturbations of relativistic stars are well-known, as they have been widely used over the last 40 years or so. Let us here very briefly recall that the fractional variations of the radial displacement and the perturbation of the pressure defined by
\begin{eqnarray}
\xi & \equiv & \frac{\Delta r}{r}  \\
\eta & \equiv & \frac{\Delta p}{p}
\end{eqnarray}
satisfy a system of two coupled first-order differential equations \cite{chanmugan, pulsating1}
\begin{eqnarray}
\xi'(r) & = & -\left(\frac{3}{r} + \frac{p'}{\zeta}\right) \xi - \frac{1}{r \Gamma} \eta \\
\eta'(r) & = & \omega^2   \left[r \left( 1 + \frac{\rho}{p} \right) e^{\lambda-\nu}\right]\xi
\nonumber
 \\
&  - &  \left[ \frac{4 p'}{p} + 8 \pi \zeta r e^{\lambda} - \frac{r (p')^2}{p \zeta}\right]\xi 
\nonumber
\\
& - &
 \left[ \frac{\rho p'}{p \zeta} + 4 \pi \zeta r e^{\lambda} \right]\eta
\end{eqnarray}
with
\begin{eqnarray}
\zeta & \equiv & p + \rho  \\
\Gamma & = & c_s^2 \: \left (1 + \frac{\rho}{p} \right) \\
c_s^2 & = & \frac{dp}{d \rho}
\end{eqnarray}
with $\Gamma$ being the relativistic adiabatic index, and $c_s^2$ being the speed of sound.

The unknown frequencies $\omega$ are determined solving the corresponding 
Sturm-Liouville eigenvalue problem imposing the following boundary conditions \cite{pulsating9}
\begin{eqnarray}
\xi(0) & = & 1 \\
\eta(0) & = & -3 \Gamma(0) \\
\frac{\eta}{\xi} \Bigl|_{r=R} & = & \left[ -4 + (1-2M/R)^{-1}  \left( -\frac{M}{R}-\frac{\omega^2 R^3}{M} \right )  \right]
\end{eqnarray}
both at the centre, $r=0$, and at the surface of the star, $r=R$.


\begin{figure*}[ht]
\centering
\includegraphics[width=0.6\textwidth]{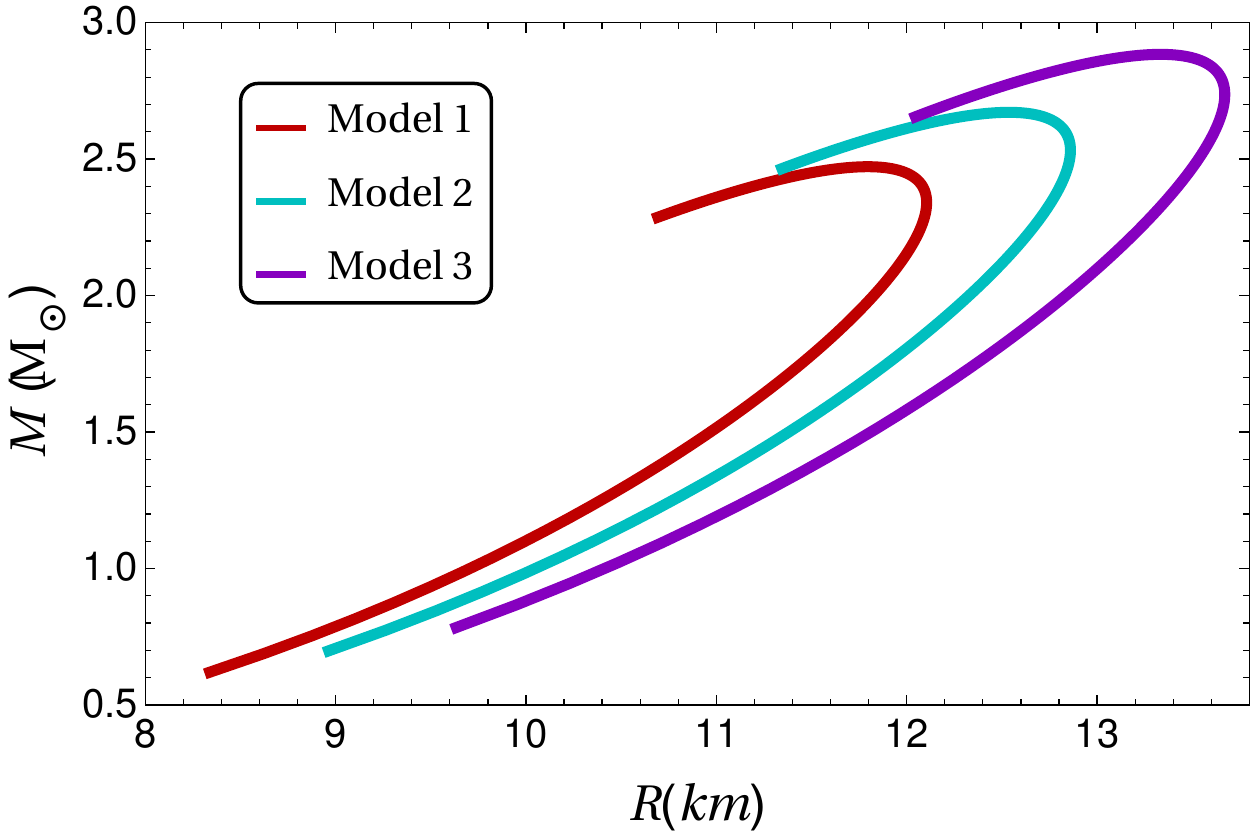}  
\caption{
Mass-to-radius relationships for the three models considered here. 
}
\label{fig:1}
\end{figure*}


\begin{figure*}[ht]
\centering
\includegraphics[width=0.6\textwidth]{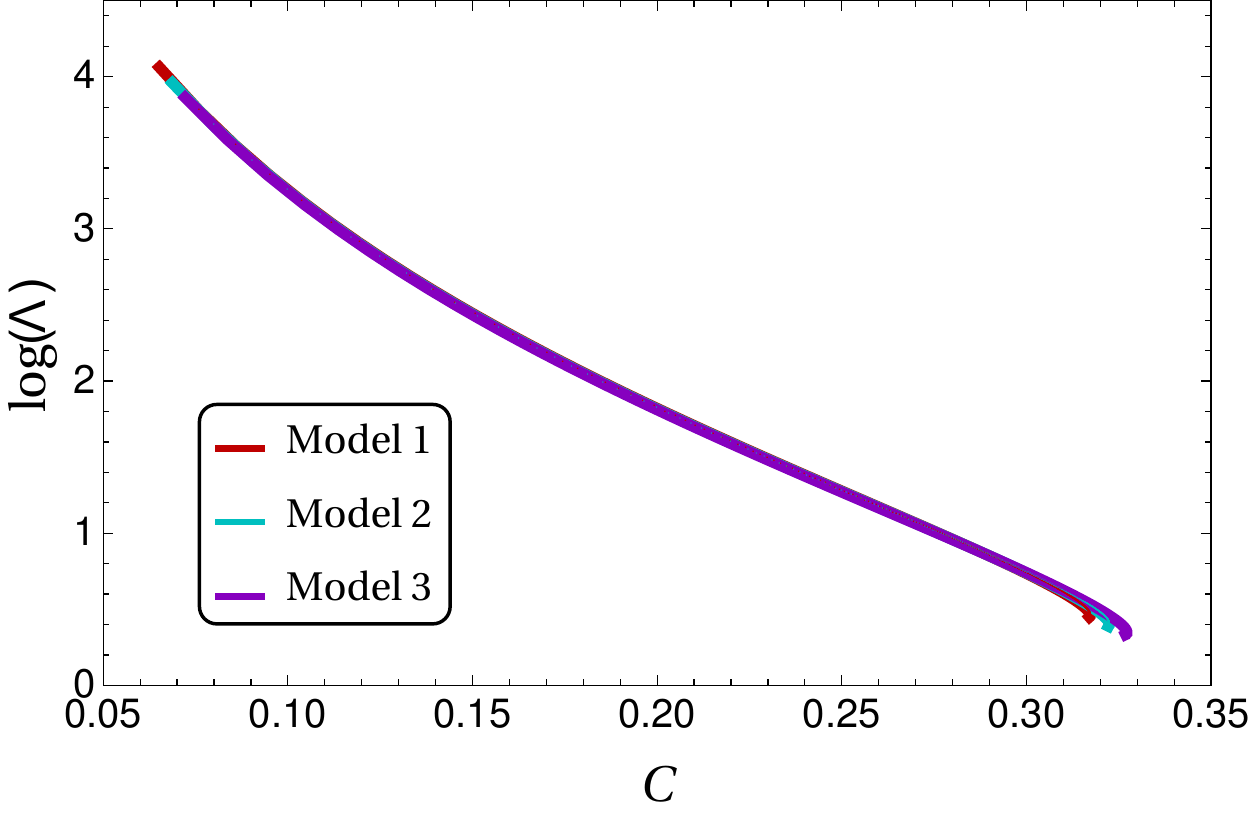}   
\caption{
Dimensionless deformability $k/C^5$ vs factor of compactness $C$ for the three models considered in the present work. 
}
\label{fig:2}
\end{figure*}


\begin{figure*}[ht]
\centering
\includegraphics[width=0.6\textwidth]{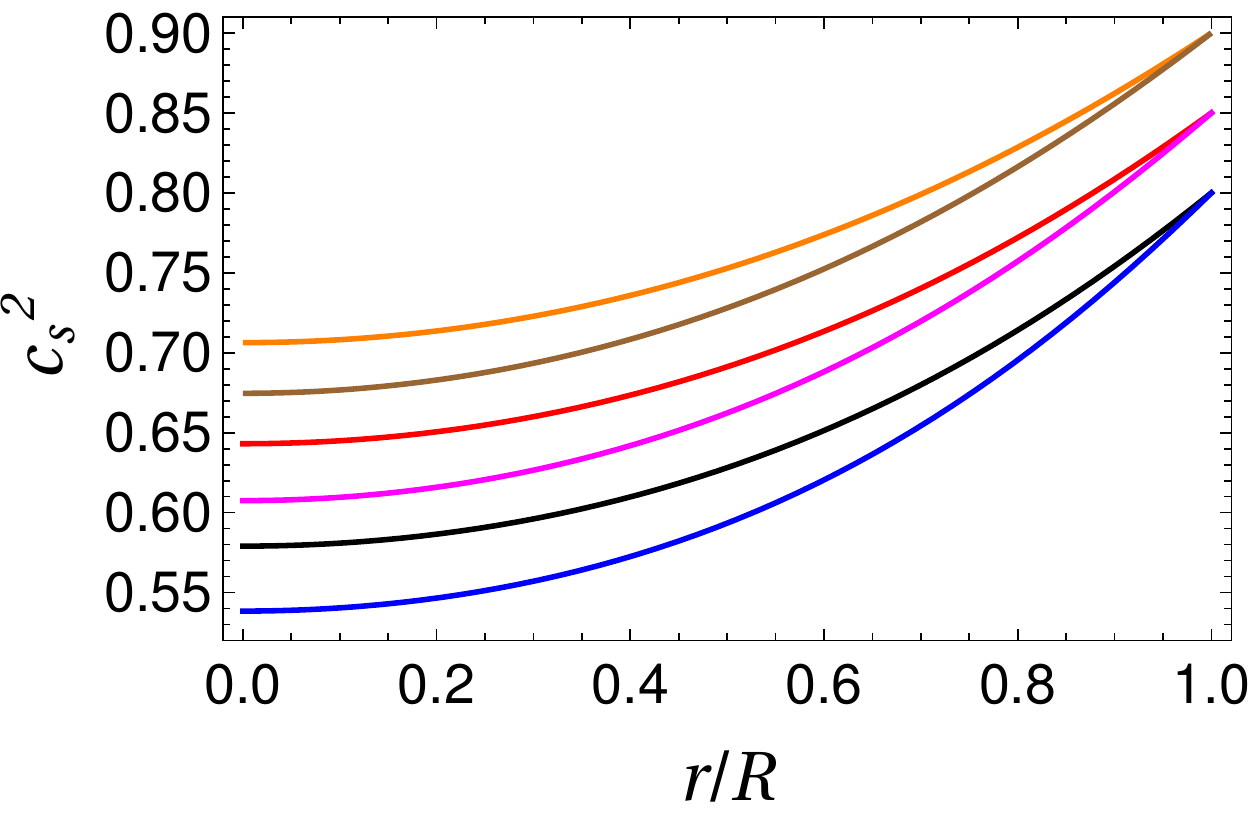}    
\caption{
Speed of sound $c_s^2$ vs dimensionless radial coordinate $r/R$ for the six fiducial stars considered here.
Shown are: $A_1$ (black), $A_2$ (blue), $B_1$ (red), $B_2$ (magenta), $C_1$ (orange), $C_2$ (brown).
}
\label{fig:3}
\end{figure*}


\begin{figure*}[ht]
\centering
\includegraphics[width=0.49\textwidth]{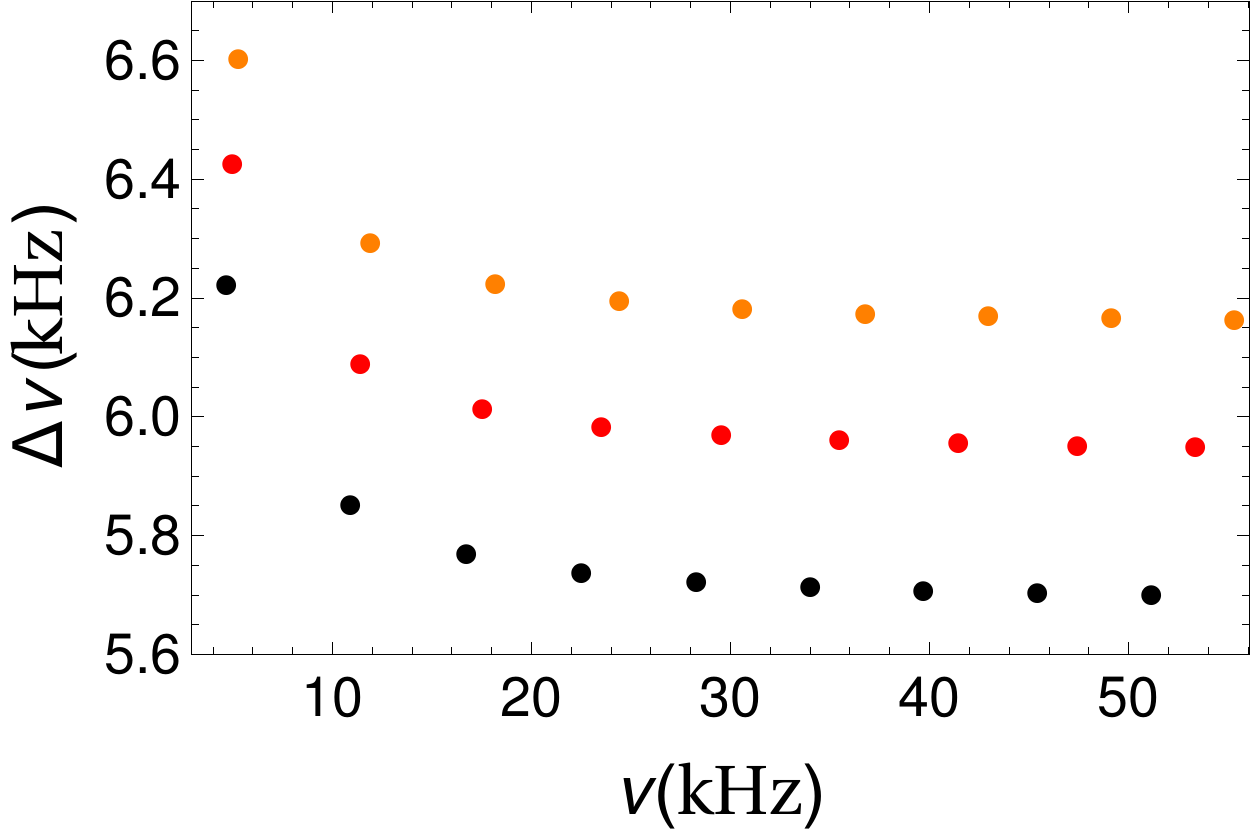}     \
\includegraphics[width=0.49\textwidth]{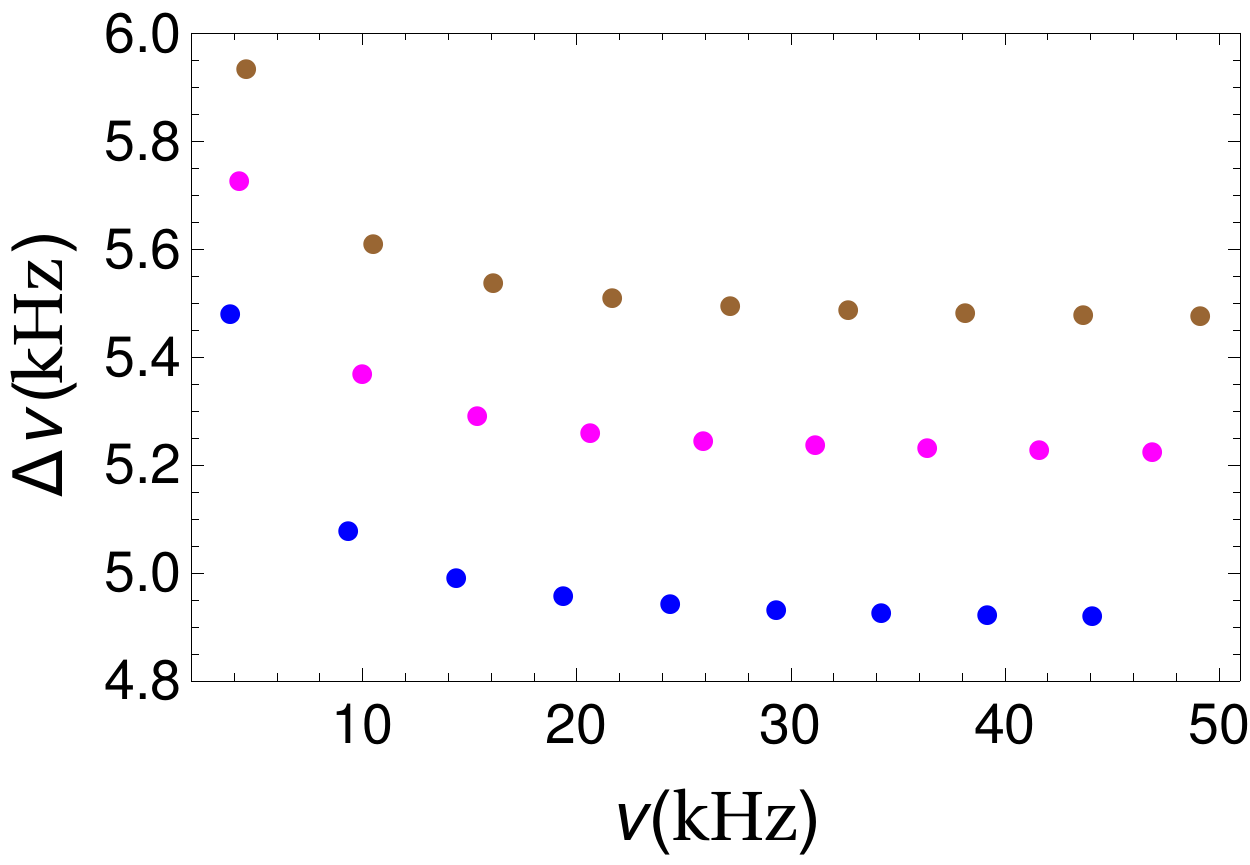}    \
\caption{
Large frequency separation $\Delta \nu_n$ (see text) vs frequencies $\nu_n$ (both in $kHz$) for the lowest 10 modes computed here.
{\bf{Left:}} Large frequency separation for the stars $A_1,B_1,C_1$ (All three have the same mass of $1.9~M_{\odot}$).
{\bf{Right:}} Same as left panel, but for the stars $A_2,B_2,C_2$ of mass $2.13~M_{\odot}$.
Shown are: $A_1$ (black), $A_2$ (blue), $B_1$ (red), $B_2$ (magenta), $C_1$ (orange), $C_2$ (brown).
}
\label{fig:4}
\end{figure*}


\section{Numerical results}

Here we present our main results, and we discuss the properties of the stars.

In Fig.~\ref{fig:1} we show the $M-R$ profiles, while in Fig.~\ref{fig:2} we show the dimensionless deformability (logarithmic plot) versus compactness for the 3 models considered in the present work. One may compare with the tidal Love numbers for other classes of stars, such as polytropes, neutron stars, strange quark stars \cite{Lattimer}, as well as boson stars, either scalar \cite{Boson} or Proca stars \cite{Proca}. We have checked numerically that the compactness of the stars takes its maximum value around $1/3$, and therefore the Buchdahl limit $C \leq 4/9$ \cite{Buchdahl:1959zz} is satisfied. 

Furthermore, we have considered six fiducial stars shown in table \ref{tab:1set}. In particular, we have considered two different masses, $1.9~M_{\odot}$ and $2.13~M_{\odot}$ from each model $A,B,C$. For those stars we show the 10 lowest radial oscillations modes (frequencies in $kHz$) in table \ref{tab:2set}. One may compare with the frequencies for neutron stars with realistic equation of state shown in table III of \cite{pulsating9}. Our results show that i) within the same model, the more massive star is characterized by lower frequencies, and ii) for stars of the same mass, the frequencies increase as we move from model $A$ towards model $C$. Finally, in Fig.~\ref{fig:4} we show the large frequency separation defined by
\begin{equation}
\Delta \nu_n \equiv \nu_{n+1} - \nu_n, \ \quad n=0,1,...8.
\end{equation}

\subsection{Stability, causality and energy conditions}

Before we finish we should check if the solutions obtained here are able to describe realistic astrophysical configurations. Regarding causality and stability criteria, we need to verify that $c_s^2 \leq 1$ and $\Gamma > 4/3$ \cite{Moustakidis:2016ndw}. The speeds of sound for all six stars are shown in Fig.~\ref{fig:3}, according to which $c_s^2$ is an increasing function of $r$ throughout the stars, and it always remains in the range $(0,1)$. Furthermore, it is easy to verify that the condition $\Gamma > 4/3$, too, is satisfied.

\smallskip

Moreover, a check regarding the energy conditions should be made. In particular, we shall now investigate weather or not the energy conditions are also fulfilled. Following \cite{Ref_Extra_1,Ref_Extra_2,Ref_Extra_3} we require that
\begin{equation}
\mbox{WEC:} \,\,\, \rho \geq 0\,, \,\,\, \rho + p \geq 0\,
\end{equation}
\begin{equation}
\mbox{NEC:} \,\,\, \rho + p  \geq  0\,
\end{equation}
\begin{equation}
\mbox{DEC:} \,\,\, \rho \geq \lvert p \rvert\,
\end{equation}
\begin{equation}
\mbox{SEC:} \,\,\, \rho + p  \geq  0\,, \,\,\, \rho + 3 p \geq 0
\end{equation} 
The (normalized) quantities $\rho/\rho_s, p/\rho_s$ exhibit the usual behaviour, namely they start from their central values at the origin, and they are monotonically decreasing functions of $r$. The pressure vanishes at the surface, whereas the energy density takes its surface value $\rho_s$. We have checked that i) both the pressure and the energy density are positive, and ii) the energy density remains larger than $p$ throughout the star. Since both quantities are positive, clearly the first, the second and the last energy conditions are fulfilled. Moreover, since $\rho$ always remains higher than $p$, the DEC is fulfilled as well. We thus conclude that the interior solutions obtained in the present work are realistic solutions within GR, and as such they are able to describe realistic astrophysical configurations.

\smallskip

As a final remark it should be stated here that in the present article we took a first step towards the investigation of static, spherically symmetric dark energy stars assuming the Chaplygin gas equation of state. It is clear that there are still several aspects that could be investigated in the future. For instance, one may study rotating stars, or stars with a net electric charge, to mention just a few. We hope to be able to address some of those interesting issues in forthcoming articles.

\section{Conclusions}

Summarizing our work, in the present article we have studied non-rotating, electrically neutral relativistic stars in general relativity. 
In particular, we have considered stars with isotropic matter assuming the generalized Chaplygin equation-of-state of the form $p = -B^2/\rho + A^2\rho$ (A and B are constants). We have computed the mass-to-radius profiles, the tidal Love numbers and the lowest radial oscillation modes of six fiducial stars. Causality, stability and energy conditions are also discussed.


\section*{Acknowlegements}

We wish to thank the anonymous reviewer for useful suggestions.
The authors G.~P. and I.~Lopes thank the Fun\-da\c c\~ao para a Ci\^encia e Tecnologia (FCT), Portugal, for the financial support to the Center for Astrophysics and Gravitation-CENTRA, Instituto Superior T\'ecnico, Universidade de Lisboa, through the Projects No.~UIDB/00099/2020 and No.~PTDC/FIS-AST/28920/2017. The author A.~R. acknowledges DI-VRIEA for financial support through Proyecto Postdoctorado 2019 VRIEA-PUCV.


\end{document}